\documentclass{article}

\def\giorno{22 April 2004}

\def\.#1{\dot #1}

\def\E{{\cal E}}
\def\F{{\cal F}}
\def\G{{\cal G}}

\def\I{{\cal I}}
\def\J{J}
\def\L{{\cal L}}

\def\R{{\bf R}}  
\def\S{{\cal S}}
\def\T{{\rm T}}

\def\X{{\cal X}}

\def\sse{\subseteq}
\def\ss{\subset}
\def \pa{\partial}

\def\=#1{\widetilde #1}
\def\.#1{\dot #1}
\def\^#1{\widehat #1}
\def\wbar#1{{\widetilde #1}}

\def\sse{\subseteq}

\def\d{{\rm d}}       

\def\({\left(}
\def\){\right)}
\def\[{\left[}
\def\]{\right]}

\def\a{\alpha}
\def\b{\beta}

\def\ga{\gamma}
\def\de{\delta}   
\def\eps{\varepsilon}
\def\phi{\varphi}
\def\la{\lambda}
\def\La{\Lambda}

\def\s{\sigma}

\def\vth{\vartheta}
\def \ep{\varepsilon}
\def \eps{\ep}
\def\Om{\Omega}
\def\phi{\varphi}
\def\Ga{\Gamma}
\def\De{\Delta}
\def\th{\theta}
\def\z{\zeta}

\def\w{\wedge}

\def\interno{\hskip 2pt \vbox{\hbox{\vbox to .18
truecm{\vfill\hbox to .25 truecm
{\hfill\hfill}\vfill}\vrule}\hrule}\hskip 2 pt}

\def\EOP{~ \hfill $\diamondsuit$} 
\def\EOD{~ \hfill $\clubsuit$}    
\def\EOR{~ \hfill $\odot$}        

\begin{document}

\title{\bf On the geometry of lambda-symmetries, and PDEs reduction}

\author{Giuseppe Gaeta\footnote{e-mail: g.gaeta@tiscali.it or giuseppe.gaeta@mat.unimi.it} \\
{\it Dipartimento di Matematica, Universit\'a di Milano} \\
{\it via Saldini 50, I--20133 Milano (Italy)} \\ {~~} \\
Paola Morando\footnote{e-mail: paola.morando@polito.it} \\
{\it Dipartimento di Matematica, Politecnico di Torino} \\
{\it Corso Duca degli Abruzzi 24, I--10129 Torino (Italy)} }

\date{Version of \giorno \ -- \ PACS: 02.20.-a ; 02.30.Jr }

\maketitle

\noindent {\bf Summary.} We give a geometrical characterization of
$\la$-prolongations of vector fields, and hence of $\la$-symmetries of ODEs. This allows an extension to the case of PDEs and systems of PDEs; in this context the central object is a horizontal one-form $\mu$, and we speak of $\mu$-prolongations of vector fields and $\mu$-symmetries of PDEs. We show that these are as good as standard symmetries in providing symmetry reduction of PDEs and systems, and explicit invariant solutions.

\section*{Introduction}

The study of differential equations was the main motivation
leading S. Lie to create what is now known as the theory of Lie
groups; symmetry methods for differential equations have received
an ever increasing attention in the last fifteen years, and by now
there is an extremely vast literature devoted to these and/or
their applications.

It was recently pointed out by Muriel and Romero \cite{MuR1} that,
beside standard symmetries, another class of transformations is
equally useful in providing symmetry reduction for scalar ordinary
differential equations (ODEs); these were christened {\it ${\cal
C}^\infty$ symmetries}, or even {\it $\la$-symmetries}, as they
depend on a smooth scalar function $\la$ (see also \cite{GMM,MuR2}
for applications of $\la$-symmetries). Soon afterwards, Pucci and
Saccomandi identified the most general class of transformations
sharing the ``useful'' properties of standard symmetries for what
concerns reduction of a scalar ODE \cite{Sac}.

In the present note, we extend the concept of $\la$-symmetries to
the case of partial differential equations (PDEs), and to systems.
In order to obtain such an extension, we found it convenient to
characterize $\la$-prolongations in $J^{(n)} M$, where $(M,\pi,B)$
is the space of dependent and independent variables seen as a
bundle over the space $B$ of independent variables, in a
geometrical way; once this characterization is obtained, it is
promptly extended from the ODEs case of $B = \R$ to the PDEs case
$B = \R^p$, and to systems.

In the scalar PDE case the transformations of interest depend on a semibasic one form $\mu = \la_i \d x^i $, the functions $\la_i$ being such to satisfy a compatibility condition. The transformations of this class leaving invariant the solution manifold for an equation $\De$ will be said to be $\mu$-symmetries, or $\La^1$-symmetries, of $\De$. In the case of systems, the form $\mu$ takes value in a Lie algebra.

We will thus be able to obtain a sound definition of
$\mu$-prolongations and $\mu$-symmetries of (systems of) PDEs.
We will also show that, in analogy with the ODE case, $\mu$-symmetries are as useful as standard symmetries in what concerns the symmetry
reduction, and the determination of invariant solutions, of (systems of) PDEs. Our approach will suffer from the same limitations as the standard PDE symmetry reduction method.

\bigskip\noindent{\bf Acknowledgements.}
We would like to thank C. Muriel and G. Saccomandi for first
explaining to us their respective works. We had since interesting
discussions on $\la$-symmetries with several colleagues and
friends; we would like in particular to thank J.F. Cari\~nena, G.
Cicogna, and G. Marmo. The work of GG was supported by {\it
``Fondazione CARIPLO per la ricerca scientifica''} under the
program {\it ``Teoria delle perturbazioni per sistemi con
simmetria''} (2000-2003).

\section{Prolongations and contact structure}

In this section we fix our notation (mainly following the usual
one in the field, see e.g. \cite{Olv1}) and recall some basic
definition \cite{Gae,Olv1,Ste,Win}.

Let us consider a space $M = B \times U$ with coordinates $x \in B
\simeq \R^p$ and $u \in U \simeq \R^q$; when setting a
differential equation in this space, we will think of the $x$ as
independent variables, and the $u$ as dependent ones.

Thus, more precisely, $M$ will be the total space of a (trivial)
linear bundle $(M,\pi,B)$ over the base space $B$, with fiber
$\pi^{-1} (x) = U$.

Given a bundle $P$, we will denote $\Ga [P]$ the set of sections
of this bundle, and by $\X [P]$ the set of vector fields in $P$.

The bundle $M$ can be {\it prolonged} to the $k$-th {\bf jet
bundle} $(J^{(k)} M,\pi_k,B)$, with $J^{(0)} M \equiv M$; the
total space of the jet bundle is also called the jet space, for
short.

The jet space $J^{(k)} M$ is naturally equipped with a canonical
{\bf contact structure} $\E$, i.e. the module generated by the set
of canonical contact one-forms
$$ \vth^a_J := \d u^a_J - u^a_{J,m} \d x^m $$
with $a=1,...,q$, $|J|=0,...,k-1$.

The contact structure in $J^{(k)} M$ defines a field of
$(p+q)$-dimensional linear spaces in $J^{(k)} M \ss \T ((J^{(k-1)}
M)$, the contact distribution, corresponding to the tangent
subspace spanned by vector fields $Y \in \X [J^{(k)} M]$
annihilated by the contact forms, i.e. such that $Y \interno
\theta = 0$ for any contact form $\theta$.
The general form of such vector fields is, as well known,
$Y = \sum \xi^i D^{(k)}_i + V$.

Here $D_i$ is the total derivative \cite{Gae,Olv1,Ste,Win} with
respect to $x^i$, $D_i^{(k)}$ its truncation to the $k$-th jet
space, and $V$ is a generic vector field in $\X [J^{(k)} M]$,
vertical for the fibration $\pi_{k,k-1} : J^{(k)} M \to J^{(k-1)}
M$ (the latter will not appear if we work with infinite-order
prolongations; it will however disappear when we deal with a given
differential equations and symmetry vector fields for it). The
operator $D_i^{(k)}$ reads
$$ D_i^{(k)} \ := (\pa / \pa x^i) \,
+ \, \sum_{a=1}^q \, \sum_{|J|=1}^{k-1} \ u^a_{J,i} (\pa / \pa
u^a_J) \ .  $$
In the following we will write, for ease of notation, simply $D_i$
instead of $D_i^{(k)}$.

A vector field $X \in \X [M]$ can be written, in the $(x,u)$ coordinates, as
$$ X \ = \ \xi^i (x,u) {\pa \over \pa x^i} \, + \, \phi^a (x,u) {\pa \over \pa u^a} \ . $$
This can be uniquely prolonged to a vector field $X^{(k)}$ in
$J^{(k)} M$ by requiring it {\it preserves the contact structure}
(the precise meaning of this will be defined in a moment). The
{\it prolongation formula} \cite{Gae,Olv1,Ste,Win} is indeed
expressing this condition.

We write a vector field in $J^{(k)} M$ as
$$ Y \ = \ X \ + \ \sum_{|J|=1}^k \,  \Psi^a_{J} {\pa \over \pa u^a_J} $$
where $X$ is as above, $J = j_1,...,j_p$ is a multiindex, and the
sum is over all multiindices of modulus $|J| = j_1 + ... + j_p$ up
to the order of the jet space. We also write $D_J$ for the total
derivative $D_{x^1}^{j_1} ... D_{x^p}^{j_p}$, and $u^a_J$ for $D_J
u^a$; moreover $u_{J,i}$ will denote $D_i u_J$.

Then $Y$ is the prolongation of $X$ if and only if the coefficients $\Psi^a_J$ satisfy the {\bf prolongation formula}
$$ \Psi^a_J \ = \ D_J \phi^a \, - \, D_J (\xi^i u^a_i) \, + \, \xi^i  D_J u^a_i \ . \eqno(1) $$
This is also recast in recursive form. We denote by $\^J = J + e_k$ the multiindex with entries $\^j_i = j_i + \delta_{ik}$, and for short $u_{J,k} := u_{J+e_k}$, $\Psi^a_{J,k} := \Psi^a_{J + e_k}$. Then (1) is equivalent to
$$ \Psi^a_{J,k} \ = \ D_k \Psi^a_J \, - \, u^a_{J,m} D_k \xi^m  \eqno(2) $$
with $\Psi^a_0 = \phi^a$ (see e.g. sect. 2.3 of \cite{Olv1}).

\bigskip

Let us now discuss the geometrical aspects of the prolongation
operation in terms of contact structures \cite{Gae,Vin,Olv1,Olv2,Ste};
these will be useful for our subsequent generalization.

Note preliminarly that for any function $f : J^{(k-1)} M \to \R$ we can write
$$ \d f \ = \ (D_i f ) \d x^i \, + \, \^\vth [f] \eqno(3) $$
where $\^\vth [f] \in \E$ is some contact form whose explicit
expression (easy to compute) is irrelevant here.

\medskip\noindent
{\bf Definition 1.} Let $Y$ be a vector field on $J^{(k)} M$. We
say that $Y$ {\bf preserves the contact structure} if  $\L_Y : \E
\to \E$. \EOD

\medskip\noindent
{\bf Proposition 1.} {\it The vector field $Y \in \X [J^{(k)} M]$,
projecting to a vector field $X \in \X [M]$ on $M$, is the
prolongation of a vector field $X \in \X[M]$ if and only if it
preserves the contact structure in $J^{(k)} M$.}

\medskip\noindent
{\bf Proof.} This is a classical result, see e.g. \cite{Vin,Olv1,Olv2,Ste}. \EOP

\medskip\noindent
{\bf Lemma 1.} {\it The vector field $Y$ preserves the contact
structure $\E$  if and only if, for any $\vth \in \E$ and any
$i=1,...p$, $ \( [D_i , Y ] \) \interno \vth \ = \ 0 $.}

\medskip\noindent
{\bf Proof.} Write $\vth^a_J$ and $Y$ as above, and note that
$D_i = \pa_i + u^a_{J,i} (\pa / \pa u^a_J)$, where of course
$\pa_i = \pa / \pa x^i$. With this notation and standard
computations,
$$ [D_i,Y] \ = \ (D_i \xi^m ) \pa_m \, + \,
(D_i \Psi^a_J - \Psi^a_{J,i} ) (\pa / \pa u^a_J ) \ ; \eqno(4) $$
hence we get $ [D_i,Y] \interno \vth^a_J = - \Psi^a_{J,i}
+ \( D_i \Psi^a_J - u^a_{J,m} (D_i \xi^m ) \)$, which
vanishes if and only if the $\Psi^a_J$ satisfy the recursive
prolongation formula (2). \EOP

\medskip\noindent
{\bf Corollary.} {\it The vector field $Y$ preserves the contact
structure $\E$  if and only if $ [D_i , Y ]  =  h_i^m D_m  +  V$
for some $h_i^m \in \La^0 (J^{(k)} M)$ and $V$ a vertical vector
field for the fibration $\pi_{k,k-1} : J^{(k)} M \to J^{(k-1)}
M$.}

\medskip\noindent
{\bf Proof.} The vector fields $D_m$ span the set of non-vertical
vector fields (for the fibration $\pi_{k,k-1}$) in the annihilator
of the contact forms. Alternatively, this follows at once from
(3), with $h_i^m = D_i \xi^m$. \EOP

\section{ODEs: $\lambda$-prolongations and $\lambda$-symmetries}

In this section we will restrict to the case of scalar ODEs, i.e. to the case where the bundle $(M,\pi,B)$ has $B=\R$ as base space and $\pi^{-1} (b) = \R$ as fiber.
We will characterize in geometrical terms, i.e. in terms of their action on the contact structure, the $\la$-prolongations
introduced by Muriel and Romero \cite{MuR1} (see also \cite{MuR2}
and \cite{GMM}), and further studied by Pucci and Saccomandi
\cite{Sac}.

We simply write $u_n$ for $D_x^n u$, and similarly for $\Psi_n$.
The standard contact forms in $J^{(k)} M$ will be $\vth_n = \d u_n
- u_{n+1} \d x$, with $n=0,...,k-1$.

We start by recalling the definition of $\la$-prolongations
and $\la$-symmetries as given by Muriel and Romero, using an
obvious notation for $x$-derivatives of the $u$.

\medskip\noindent
{\bf Definition 2.} Let $X = \xi (\pa / \pa x) + \phi (\pa / \pa
u)$ be a vector field on $M$, and  $Y = X + \sum_{n=1}^k \Psi_n
(\pa / \pa u_n)$ a vector field on $J^{(k)} M$. Let $\la : J^{(1)}
M \to \R$ be a smooth function. We say that $Y$ is the {\it
$\la$-prolongation} of $X$ if its coefficients satisfy the {\bf
$\la$-prolongation formula}
$$ \Psi_{n+1} \ = \ [(D_x + \la) \Psi_n] \, - \, u_{n+1} [(D_x + \la) \xi ] \eqno(5) $$
for all $n=0,...,k-1$. \EOD
\medskip

\medskip\noindent
{\bf Definition 3.} Let $\De$ be a $k$-th order ODE for $u = u
(x)$, $u \in U = \R$, and let $(M = U \times B, \pi, B)$ be the
corresponding variables bundle. Let the vector field $Y$ in
$J^{(k)} M$ be the $\la$-prolongation of the Lie-point vector
field $X$ in $M$. Then we say that $X$ is a {\bf $\la$-symmetry}
of $\De$ if and only if $Y$ is tangent to the solution manifold
$S_{\De}$, i.e. iff there is a smooth function $\Phi$ on $J^{(k)}
M$ such that $Y(\De) = \Phi  \De $. \EOD
\medskip

\medskip\noindent
{\bf Remark 1.} We stress that in this note we take $\la : J^{(1)}
M \to \R$, which guarantees that the $\la$ prolongation of a
Lie-point vector field in $M$ is a proper vector field in each
$J^{(n)} M$. One could also consider $\la : J^{(r)} M \to \R$,
obtaining obvious generalizations of the results given here. In
this case the $\la$-prolongations of $X$ would be generalized
vector fields in each $J^{(n)} M$ with $n>0$ even if $X$ is a
Lie-point vector field. The same applies to the
$\mu$-prolongations to be considered in later sections. \EOR
\medskip

We will not discuss here the relevance of $\la$-symmetries,
referring to \cite{MuR1,Sac}; we just recall that they are as
useful as standard ones in that one can perform symmetry reduction
to the same extent as for standard symmetries \cite{MuR1}.

The basic property of $\la$-prolongations behind this feature was
clearly pointed out by Pucci and Saccomandi \cite{Sac}, and can be
expressed in terms of the characteristics of the vector fields $Y$
which are $\la$-prolongations of $X$.

\bigskip

Given the vector bundle $(M,\pi,B)$, we choose a distinguished
smooth real function $\la (x,u,u_x) : J^{(1)} M \to \R$. We note
for later discussion that to this is associated a semibasic
one-form $\mu \in \La^1 (J^{(1)} M)$, i.e. the one-form $ \mu =
\la (x,u,u_x) \d x$

\medskip\noindent
{\bf Definition 4.} Let $Y$ be a vector field on the contact
manifold $(J^{(k)} M,\E)$, and $\la \in \La^0 (J^{(1)} M)$ a
smooth function on $M$. We say that $Y$ {\bf $\la$-preserves the
contact structure} if, for any contact one-form $\th \in \E$,
$$ \L_Y (\th ) \ + \ (Y \interno \th ) \, \la \d x \ = \ \^\th \eqno(6) $$
for some contact one-form $\^\th \in \E$. \EOD
\medskip

\medskip\noindent
{\bf Theorem 1.} {\it Let $(M,\pi,B)$ be a bundle over the real
line $B = \R$ with fiber $\pi^{-1} (x) = \R$, and let $\E$ be the
standard contact structure in $J^{(k)} M$. Let $Y$ be a vector
field on the jet space $J^{(k)} M$, which projects to a vector
field $X$ on $M$. Then $Y$ is the $\la$-prolongation of $X$ if and
only if it $\la$-preserves the contact structure.}

\medskip\noindent
{\bf Proof.} We write a general vector field on $J^{(k)} M$ as $Y
= \xi \pa_x + \sum_{m=0}^k \Psi_m (\pa / \pa u_m )$; as the
contact forms are $\vth_n = \d u_n - u_{n+1} \d x$
($n=0,...,k-1$), we have by explicit computation
$$ \L_Y (\vth_n) + (Y \interno \vth_n) \la \d x  =
\[ - \Psi_{n+1} + D_x \Psi_n - u_{n+1} D_x \xi + \la ( \Psi_n - u_{n+1} \xi ) \] \d x
+ \^\th $$
with $\^\th$ a contact form. Thus (6) is satisfied if and only if the $\Psi_n$ satisfy the $\la$-prolongation formula (5). \EOP
\bigskip

We can also provide an alternative characterization of $\la$-prolonged vector fields, similarly to what we did for standard prolongations in lemma 1.

\medskip\noindent
{\bf Lemma 2.} {\it Let $Y$ be a vector field on the jet space
$J^{(k)} M$, with $(M,\pi,B)$ a vector bundle over the real line
$B = \R$, and let $\E$ be the standard contact structure in
$J^{(k)} M$. Then $Y$ is the $\la$-prolongation of a vector field
$X$ on $M$ if and only if, for any $\vth \in \E$,
$$  [ D_x , Y ] \interno \vth \ = \ \la (Y \interno \vth ) \ . \eqno(7) $$}

\medskip\noindent
{\bf Proof.} Looking at the proof of lemma 1, $[D_x,Y] \interno \vth$ is given by (4) specialized to the case $p=1$: with the obvious notation $u_n := D_x^n u$ (and similarly for $\Psi_n$) we have $ [D_x,Y] = - \Psi_{n+1} + ( D_x \Psi_n - u_{n+1} D_x \xi )$; on the other hand, it is easy to check that $Y \interno \vth_n \ = \ \Psi_n - u_{n+1} \xi$. Thus eq. (7) is equivalent to $\Psi_{n+1} = [( D_x + \la ) \Psi_n] - u_{n+1} [(D_x + \la ) \xi ]$, i.e. to the $\la$-prolongation formula (5). \EOP
\bigskip

\medskip\noindent
{\bf Corollary.} {\it In the hypotheses of lemma 2, $Y$ is the
$\la$-prolongation of a vector field $X$ on $M$, if and only if $
[ D_x , Y ]  = \la Y + h D_x + V $ with $\la , h$ scalar functions
on $J^{(1)} M$ and $V$ a vertical vector field for the fibration
$\pi_{k,k-1} : J^{(k)} M \to J^{(k-1)} M$.}

\medskip\noindent
{\bf Remark 2.} Theorem 1 shows that our geometrical formulation,
i.e. definition 4, is equivalent to the standard (analytical) one,
i.e.  definition 2. The advantage of our formulation is twofold:
we have a geometrical characterization of $\la$-prolongations
($\la$-symmetries), and moreover this is readily extended from the
ODEs to the PDEs case. As we discuss later on, we can moreover
generalize the standard symmetry reduction method for PDEs to an
analogous $\la$-symmetry reduction. We will also show that this
definition and the reduction procedure extend to systems of PDEs.
\EOR

\section{PDEs: $\mu$-prolongations and $\mu$-symmetries}

In this section we extend our approach to $\la$-prolongations and
$\la$-symmetries to the case of scalar PDEs ($p$ independent
variables); the case of PDE systems will be dealt with in section
5 below.

The role of the scalar function $\la$ will now be played by an
array of $p$ smooth functions $\la_i : J^{(1)} M \to \R$ (remark 1
holds also in this context), which will be the components of a
semibasic form $\mu \in \La^1 (J^{(1)} M)$. The only additional
ingredient required in the multi-dimensional (PDE) case is a
compatibility condition between the semibasic form $\mu$ and the
contact structure -- this is eq.(10) below -- automatically
satisfied in the ODE case.

Actually, our formulation of $\la$-prolongations in the ODE case
was such that the results, and even their proofs, are the same
also in the PDE case -- except of course for the appearance of new variables.

In view of our geometric approach it is convenient to focus on the form $\mu$ rather than on the $q$-ple of smooth functions $\la_i$. We will thus call the analogue of $\la$-prolongations and $\la$-symmetries in the PDE frame, $\mu$-prolongations and $\mu$-symmetries.

We equip $(J^{(1)} M,\pi,B)$ with a distinguished semibasic
one-form $\mu$,
$$ \mu \ = \ \la_i  \, \d x^i \ . \eqno(8) $$
We require that $\mu$ is {\it compatible} with the contact
structure defined in $J^{(k)} M$, for $k \ge 2$, in the sense that
$$ \d \mu \in \J(\E) \ , \eqno(9) $$
where $\J(\E)$ is the Cartan ideal generated by $\E$ (we recall
that a two-form $\a$ is in $\J (\E)$ if and only if $\a = \rho^J
\w \vth_J$ for some one-forms $\rho^J$).

It should be noted that this condition does not appear when we
deal with first order equations, i.e. with first order
$\mu$-prolongations. We note also that for $p=1$ eq.(9) is automatically satisfied:
indeed, $\d \mu = (\pa \la / \pa u) \d u \w \d x + (\pa \la / \pa
u_x) \d u_x \w \d x = (\pa \la / \pa u) \vth_0 \w \d x + (\pa \la
/ \pa u_x) \vth_1 \w \d x$.

\medskip\noindent
{\bf Lemma 3.} {\it Condition (9) is equivalent to
$$ D_i \la_j - D_j \la_i \ = \ 0 \ . \eqno(10) $$
This is in turn equivalent to the condition that the operators $\nabla_i := D_i + \la_i$ commute, $[\nabla_i , \nabla_j ] = 0$.}

\medskip\noindent
{\bf Proof.} As $\la_i$ is a function on $J^{(1)} M$, we have $\d
\mu = (\pa \la_j / \pa x^i ) \d x^i \w \d x^j  +  (\pa \la_j / \pa
u ) \d u \w \d x^j + (\pa \la_j / \pa u_i ) \d u_i \w \d x^j$,
i.e. $$ \begin{array}{rl} \d \mu =& [ (\pa \la_j / \pa x^i ) + u_i
(\pa \la_j / \pa u ) + u_{ik} (\pa \la_j / \pa u_k ) ] \d x^i \w
\d x^j \, + \\ & + \, (\pa \la_j / \pa u ) \vth_0 \w \d x^j \, +
\, (\pa \la_j / \pa u_i ) \vth_i \w \d x^j \ . \end{array} $$ The
two latter terms are of course in $\J(\E)$, while no form $\d x^i
\w \d x^j$ belongs to $\J(\E)$; thus, (9) is satisfied if and only
if the coefficients of all these terms vanish. This condition is
precisely (10). (Note this extends to the case where $\mu$ is
semibasic for $(J^{(n)} , \pi_n , B)$, see remark 1.) The equivalence of this with $[\nabla_i,\nabla_j] = 0$ follows from the definition of $\nabla_i$. \EOP
\medskip

\medskip\noindent
{\bf Definition 5.} Let $Y$ be a vector field on the contact
manifold $(J^{(k)} M,\E)$, and $\mu$ a semibasic form on $M$
compatible with $\E$. We say that $Y$ {\bf $\mu$-preserves the
contact structure} if, for any $\th \in \E$, there is a form
$\^\th \in \E$ such that
$$ \L_Y (\th ) \ + \ (Y \interno \th ) \mu \ = \ \^\th \ . \eqno(11) $$
\EOD

\medskip\noindent
{\bf Definition 6.} A vector field $Y$ in $J^{(k)} M$ which
projects to $X$ in $M$ and which $\mu$-preserves the contact
structure is said to be the  {\bf $\mu$-prolongation} of order
$k$, or the $k$-th $\mu$-prolongation, of $X$. \EOD

\medskip\noindent
{\bf Theorem 2.} {\it Let $Y$ be a vector field on the jet space
$J^{(k)} M$, with $(M,\pi,B)$ a vector bundle over $B = \R^p$,
written in coordinates as
$$ Y \ = \ X + \sum_{|J|=1}^k \Psi_J {\pa \over \pa u_J} \ , $$
with $X = \xi^i (\pa / \pa x^i ) + \phi (\pa / \pa u)$ a vector
field on $M$. Let $\E$ be the standard contact structure in
$J^{(k)} M$, and $\mu = \la_i \d x^i$ a semibasic one-form on
$(J^{(1)} M,\pi,B)$, compatible with $\E$. Then $Y$ is the
$\mu$-prolongation of $X$ if and only if its  coefficients (with
$\Psi_0 = \phi$) satisfy the {\bf $\mu$-prolongation formula}
$$ \Psi_{J,i} \ = \ (D_i + \la_i ) \Psi_J \, - \, u_{J,m} \, (D_i + \la_i ) \xi^m \ . \eqno(12) $$}

\medskip\noindent
{\bf Proof.} The standard contact forms in $J^{(k)} M$ are $\vth_J
= \d u_J - u_{J,i} \d x^i$, with $|J|=0,...,k-1$. Thus, as already
computed in the proof of proposition (1), $\L_Y (\vth_J) = ( -
\Psi_{J,i} + D_i \Psi_J - u_{J,m} D_i \xi^m ) \d x^i + \Theta$
with $\Theta$ a contact form. On the other hand, it is easy to
compute that $Y \interno \vth_J = \Psi_J - u_{J,m} \xi^m$.
Therefore,
$$ \begin{array}{l}
\L_Y (\vth_J) + (Y \interno \vth_J ) \mu \ = \\
\ \ \ = \ \[ ( - \Psi_{J,i} + D_i \Psi_J - u_{J,m} D_i \xi^m ) + \la_i (\Psi_J - u_{J,m} \xi^m) \] \d x^i \, + \, \Theta \ .
\end{array} $$
This is a contact form if and only if the coefficients of all the
$\d x^i$ vanish, i.e. if and only if (12) is satisfied. \EOP

\medskip\noindent
{\bf Remark 3.} Condition (9) arises from the following: consider
the multiindices $J = (j_1,...,j_p)$ and $L = (\ell_1,...,
\ell_p)$ with $\ell_s = j_s + \de_{i,s} + \de_{k,s}$; the
coefficient $\Psi_{L}$ can be obtained from $\Psi_J$ by applying
twice formula (12), but we can proceed in two different ways, i.e.
pass first from $\Psi_J$ to $\Psi_{J,i}$ and then to $\Psi_L$, or
pass first from $\Psi_J$ to $\Psi_{J,k}$ and then to $\Psi_L$.
Needless to say, the result must be the same in the two cases, and
this is the {\bf compatibility condition} for the $\la_i$. By
explicit computation this is just (10), equivalent to (9) by lemma
3. \EOR
\medskip

As for standard and $\la$-prolongations, $\mu$-prolongations have
a specific behaviour for what concerns their commutation with the
total derivatives $D_i$.

\medskip\noindent
{\bf Lemma 4.} {\it If $Y$ is the $\mu$-prolongation of a
Lie-point vector field $X$, with $\mu = \la_i \d x^i$, then for
any contact form $\vth$,
$$ [D_i,Y] \interno \vth \ = \ \la_i \, (Y \interno \vth) \ . \eqno(13) $$}

\medskip\noindent
{\bf Proof.} In the proof of lemma 1 we have computed $ [D_i , Y]
\interno \vth_J = - \Psi_{J,i} + D_i \Psi_J - u_{J,m} D_i \xi^m$;
needless to say, $Y \interno \vth_J = \Psi_J - u_{J,m} \xi^m$ and
thus (13) is equivalent to the $\mu$-prolongation formula (12).
\EOP

\medskip\noindent
{\bf Corollary.} {\it In the hypotheses of lemma 4, $Y$ is the
$\mu$-prolongation of a vector field $X$ on $M$, if and only if $
[ D_i , Y ] \ = \ \la_i Y  + h_i^m D_m + V $ with $\la_i , h_i^m$
scalar functions on $J^{(1)} M$ and $V$ a vertical vector field
for the fibration of $J^{(k)} M$ over $J^{(k-1)} M$.}
\medskip

It is quite remarkable that a simple relation exists between the
$\mu$-prolongation of a vector field and its ordinary
prolongation. In order to discuss this relation,
we write $X = \xi^i (\pa / \pa x^i) + \phi (\pa / \pa u)$ for the
vector field in $M$, and denote its ordinary prolongations as
$X^{(k)} = X + \wbar{\Psi}_J (\pa / \pa u_J)$, while its
$\mu$-prolongations are denoted as $ Y = X + \Psi_J (\pa / \pa
u_J)$. The form $\mu$ is written, as usual, $\mu = \la_i \d x^i$,
and of course $\Psi_J = \wbar{\Psi}_J$ when all the $\la_i$ (or at
least all those for $i$ such that $j_i \not= 0$) vanish.

As well known \cite{Gae,Olv1,Ste,Win}, the vector field $X$ can be cast in evolutionary form as $X_Q := Q (\pa / \pa u)$, with $Q := \phi - u_i \xi^i$.

The equations $D_J Q = 0$, with $|J|=0,...,k-1$ identify the
$X$-invariant space $\I_X \ss J^{(k)} M$. We denote by $\F$ the
module over ${\cal C}^\infty (J^{(k)} M)$ generated by the $D_J
Q$, i.e. the  set of functions $F$ which can be written as $F =
c^J D_J Q$ for some smooth functions $c^J : J^{(k)} M \to \R$, and
by $\F^{(m)} \sse \F \equiv \F^{(k)}$ those which depend only on
variables $(x,u^{(m)})$, $m \le k$. Needless to say, $D_i :
\F^{(m-1)} \to \F^{(m)}$.

\medskip\noindent
{\bf Theorem 3.} {\it Let $X,Y,\mu$ be as above. Write $\Psi_J$ as
$ \Psi_J = \wbar{\Psi}_J + F_J$.
Then the functions $F_J$ satisfy the recursion relation (with $F_0
= 0$)
$$ F_{J,i} \ = \ ( D_i + \la_i ) F_J \, + \, \la_i D_J Q \ . \eqno(14) $$
}

\medskip\noindent
{\bf Proof.} In order to show that the statement of the theorem
holds at all orders, we proceed recursively: we suppose (14) holds
for all $|J| < h$, and wish to prove that it holds also for
$|J|=h$. Any $\^J$ of order $h$ can be written as $J + e_i$ for
some $i$ and some $J$ of order $h-1$; formula (14) holds for
$\Psi_J$. Thus, by the $\mu$-prolongation formula,
$$ \begin{array}{rl}
\Psi_{J,i} \ =& \ (D_i + \la_i ) \wbar{\Psi}_J - u_{J,m} (D_i +
\la_i ) \xi^m + (D_i + \la_i ) F_J \\
=& \ [ D_i \wbar{\Psi}_J - u_{J,m} D_i \xi^m ] + \la_i [
\wbar{\Psi}_J - u_{J,m} \xi^m ] + (D_i + \la_i ) F_J \ .
\end{array} $$
The first term is just $\wbar{\Psi}_{J,i}$, and the
last is already in the form appearing in (14); so we have to look
only at the second one.

Take an $s$ such that $j_s \not= 0$ in the multiindex $J$, and
write $K = J - e_s$. Then, using the standard prolongation
formula,
$$ \wbar{\Psi}_J - u_{J,m} \xi^m  = [ D_s \wbar{\Psi}_{K} -
u_{K,m} D_s \xi^m ] - ( D_s u_{K,m} ) \xi^m = D_s  \(
\wbar{\Psi}_{K} - u_{K,m} \xi^m \) \ . $$ We can then repeat the
procedure on any index $q$ such that $K=J - e_s$ has a nonzero $q$
entry, and so on. In the end, recalling that $\Psi_0 = \phi$, we
have
$$ \wbar{\Psi}_J - u_{J,m} \xi^m \ = \ D_J \( \phi - u_m \xi^m \) \ = \ D_J Q \ ; $$
this also follows from the formula for prolongation of the
evolutionary representative of $X$. Going back to our computation,
we have thus shown that
$$ \Psi_{J,i} \ = \ \wbar{\Psi}_{J,i} \ + \ \la_i D_J Q \ + \ (D_i + \la_i ) F_J \ . $$
This shows that if (14) is satisfied at order $h-1$, it is also
satisfied at order $h$.

It is easy to check that (14) holds at order one, i.e. for $|J| =
0$: indeed, by the $\mu$-prolongation formula (12) and the
ordinary prolongation formula (1),
$$ \begin{array}{rl}
\Psi_i & = \ (D_i + \la_i ) \phi - u_m (D_i + \la_i ) \xi^m \\
& = \ \( D_i \phi - u_m D_i \xi^m \) \, + \, \la_i \( \phi - u_m \xi^m \) \\
& = \ \wbar{\Psi}_i + \la_i Q \ . \end{array} $$
We conclude that (14) holds at all orders. \EOP
\medskip

This theorem provides an economic way of computing
$\mu$-prolongations of $X$ if we already know its ordinary
prolongations. Theorem 3 also has a rather obvious consequence,
which will be relevant in the following.

\medskip\noindent
{\bf Lemma 5.} {\it Let $X$ be a vector field on $M$, $\E$ the
standard contact structure on $J^{(k)} M$, and $\mu$ any semibasic
form on $M$ compatible with $\E$. Then: {\rm (i)} the
$\mu$-prolongation $Y$ of $X$ coincides with the ordinary
prolongation $X^{(k)}$ on the invariant space $\I_X$; {\rm (ii)}
the space $\I_X \ss J^{(k)} M$ is invariant under the
$\mu$-prolongations of $X$, for any semibasic form $\mu$
compatible with $\E$.}

\medskip\noindent
{\bf Proof.} By definitions, any function $F \in \F$ vanishes
identically on $\I_X$. Thus (14) guarantees that $\Psi_J =
\wbar{\Psi}_J$ on $\I_X$, i.e. proves point {\it (i)}.

As for {\it (ii)}, this is a known property of standard
prolongations, easily checked by using the evolutionary
representative of $X$, $X_Q := Q (\pa / \pa u)$. Its prolongation
is $X_Q^{(k)} = (D_J Q) (\pa / \pa u_J)$, where the sum is over
all multiindices with $|J| \le k$, and $X^{(k)} = X_Q^{(k)} +
\xi^i D_i $. Thus, $X^{(k)}$ reduces to $W = \xi^i D_i$ on $\I_X$;
and $W$ is obviously tangent to $\I_X$. \EOP
\medskip

Finally, we define $\mu$-symmetries of a PDE as Lie-point vector
fields whose $\mu$-prolongation is a symmetry of the equation.

\medskip\noindent
{\bf Definition 7.} Let $X$ be a vector field on $M$, and let $Y
\in \X [J^{(k)} M]$ be its $\mu$-prolongation of order $k$. Let
$\De$ be a differential equation of order $k$ in $M$, $\De :=
F(x,u^{(k)} ) = 0$, and $\S \ss  J^{(k)} M$ be the solution
manifold for $\De$. If $Y : \S \to \T \S$, we say that $X$ is a
{\bf $\mu$-symmetry} for $\De$. If $Y$ leaves invariant each level
manifold for $F$, we say that $X$ is a {\bf strong $\mu$-symmetry}
for $\De$. \EOD

\medskip\noindent
{\bf Remark 4.} Note that if we look for $\mu$-symmetries of a
given equation $\De$, we can accept forms $\mu$ which do not
satisfy (9) on the whole jet space $J^{(n)} M$, but only on the
solution submanifold $S_\De \ss J^{(n)} M$.  \EOR

\medskip\noindent
{\bf Remark 5.} Given a form $\mu = \la_i \d x^i$, we consider
exponential vector fields
$$ X \ = \ e^{\int \mu} \cdot \ X_0 $$
where $X_0$ is a vector field on $M$; note that if $\mu = (D_i P)
\d x^i$, in which case (10) is automatically satisfied, then $X =
e^P X_0$; in general $X$ is a formal expression. For a general
$\mu$, consider an equation $\De$ such that (10) is satisfied on
$S_\De$, see the remark above. Then we have the following result
(see \cite{CGM} for a proof and extensions): $X$ is a (in general,
nonlocal) symmetry for $\De$ if and only if $X_0$ is a
$\mu$-symmetry for $\De$. This extends a result by Muriel and
Romero \cite{MuR1}. \EOR
\medskip

The relevant point is that {\bf $\mu$-symmetries can be used to
obtain group-invariant solutions}, i.e. one can introduce
$\mu$-symmetry reductions of PDEs and obtain invariant solutions
to the original PDE from these, by the same method as for standard
symmetries.

Note that in this way we parallel again the ODE case, where it was
proven by Muriel and Romero and by Pucci and Saccomandi that
$\la$-symmetries are as good as standard ones for reduction of the
equation.

\section{The $\mu$-symmetry reduction method for PDEs}

As well known, symmetry reduction for PDEs is conceptually
different from symmetry reduction for ODEs: while in the latter
case it yields a reduced equation whose solutions provide,
together with an integration, the most general solution to the
original ODE, in the PDE case the reduced equation provides only
the symmetry-invariant solutions to the original PDE.

\subsection{The PDE reduction method}

In this subsection we briefly recall (using the notation
introduced so far) symmetry reduction for scalar PDEs in the case
of standard symmetries; this is discussed in detail in a number of
textbooks and research papers, see e.g. \cite{Gae,Olv1,Ste,Win}.
We will just discuss reduction under a single vector field, rather
than a general (i.e. higher dimensional) Lie algebra.

Consider a PDE of order $k$ $\De$, which we may think in the form
$F (x,u^{(k)}) = 0 $ with $F : J^{(k)} M \to \R$ a smooth scalar
function. Let the Lie-point vector field $X$ in $M$, with
prolongation $X^{(k)}$ in $J^{(k)} M$, be a (standard) symmetry
for $\De$. Then we proceed as follows, following Olver. (For more
details, see e.g. the discussion in chapter 3 of \cite{Olv1}).

First of all we pass to symmetry-adapted coordinates in $M$. In
practice, we have to determine a set of $p$ independent invariants
for $X$ in $M$, which we will denote as $(y^1,...,y^{p-1},v)$:
these will be our $X$-invariant coordinates, and essentially
identify the $G$-orbits, while the remaining coordinate $\s$ will
be acted upon by $G$. In other words, $G$-orbits will correspond
to fixed value of $(y,v)$ coordinates and to $\s$ taking values in
a certain subset of the real line (thus $(y,v)$ are coordinates on
the {\it orbit space} $\Om = M/G$, see \cite{Olv1})

The invariants will be given by some functions $y^i = \eta^i
(x,u)$ ($i=1,...,p-1$) and $v = \zeta (x,u)$ of $x^1,...,x^p$ and
$u$. If $X$ acts transversally, we can invert these for $x$ and
$u$ as functions of $(y,v;\s)$, i.e. write $ x^i = \chi^i
(y,v;\s)$ ($i=1,...,p$) and $u = \b (y,v;\s)$.

If now we decide to see the $(y;\s)$ as independent variables and
the $v$ as the dependent one, we can use the chain rule to express
$x$-derivatives of $u$ as $\s$ and $y$-derivatives of
$v$\footnote{It is maybe worth recalling that this computation can
be described also in a slightly different, but equivalent, way:
that is, we write $\d u = u_i \d x^i$ on the one hand, and $\d u =
\d [\b (y,v;\s) ]$ on the other. We then expand the latter as $\d
u = \b_j \d y^j + \b_v \d v + \b_\s \d \s$, where of course $\b_j
= \pa \b / \pa y^j$, and substitute for $\d v$ as $\d v = v_j \d
y^j + v_\s \d \s$. Comparing the two expressions for $\d u$, we
obtain the expression for $u_i$ in terms of $v_j$ and $v_\s$.}.
Using these, we can finally write $\De$ in terms of the $(y,v;\s)$
coordinates and derivatives of $v$ in the $y$ and $\s$; this will
turn out to be an equation which, when subject to the side
condition $\pa v / \pa \s = 0$, is independent of $\s$. The
condition $\pa v / \pa \s = 0$ expresses the fact that the
solutions are required to be invariant under $X$, i.e. the
equation obtained in this way represents the restriction of $\De$
to the space of $G$-invariant functions, and therefore it is
sometimes also denoted as $\De /G$.

Suppose we are able to determine some solution $v = \Phi (y)$ to
the reduced equation; we can write this in terms of the $(x,u)$
coordinates as $ \zeta (x,u) = \Phi [ \eta (x,u) ] $, which yields
implicitly $u = f (x)$: this is the corresponding $X$-invariant
solution to the original equation $\De$ in the original
coordinates.

\medskip\noindent
{\bf Remark 6.} The symmetry reduction method for PDEs can also be
seen in a slightly different way: if we look for $X$-invariant
solution $u=f(x)$ to $\De$, we determine the characteristic $Q =
\phi - u_i \xi^i$ of the vector field $X$, and supplement $\De$
with the equations $E_J := D_J Q = 0$ with $|J|=0,...,k-1$. The
equation $E_0$ requires that the evolutionary representative $X_Q
= Q (\pa / \pa u)$ vanish on $\ga_f$, i.e. that $u$ is
$X$-invariant, and all the equations with $|J| >0$ are just
differential consequences of this. The $X$-invariant solutions to
$\De$ are in one to one correspondence with the solutions to the
system $\De_{(X)} := \{ \De ; E_J \}$. See e.g. \cite{Win} for
details, and for how this approach is used in a more general
context. \EOR

\medskip\noindent
{\bf Remark 7.} We stress that the standard method discussed here
applies under a nondegeneracy (transversality) condition,
guaranteeing a certain Jacobian admits an inverse. When this is
not the case -- as it happens in a number of physically relevant
cases -- the treatment should go through the approach developed by
Anderson, Fels and Torre \cite{AFT}. See also \cite{GTW} for the
case of partial transversality. We also stress that this method is
justified only if the (possibly, only local) one-parameter group
$G$ generated by $X$ has regular action in $M$, i.e. the
$G$-orbits are regular embedded submanifolds of $M$ \cite{Dui}. In
the following we tacitly assume both conditions mentioned here are
satisfied. \EOR

\subsection{On the justification of the method}

The method described above is rigorously justified in chapter 3 of
\cite{Olv1}, to which we refer for details. In this subsection we
just recall what is the key step in the proof, as we will have to
prove a similar property also holds for $\mu$-prolongations in
order to justify the extension of this method to $\mu$-symmetries.

We recall that a function $u = f (x^1,...,x^p)$ corresponds to a
section $\ga_f \in \Ga [M]$, the space of sections for the bundle
$(M,\pi,B)$, i.e. $\ga_f = \{ (x,u) : u = f (x) \}$. This is
uniquely prolonged to a section $\ga_f^{(k)} \in \Ga [ J^{(k)}
M]$; $\ga_f^{(k)}$ is the unique lift of the curve $\ga_f$ in $M$
to a curve in $J^{(k)} M$ which $(i)$ projects down to $\ga_f$ in
$M$, and $(ii)$ is everywhere tangent to the field of contact
linear spaces.

When we act with a vector field $X = \xi^i (\pa / \pa x^i) + \phi
(\pa / \pa u)$ on $M$, at the infinitesimal level, the section
$\ga_f$ is mapped into $\ga_{\^f}$ with
$$ {\^f} (x) \ = \ f (x) \ + \ \eps [ \phi (x,u) -
\xi^i (x,u) \pa_i f (x) ]_{u=f(x)} \ + \ o (\eps) \ . $$

Thus a function $u=f(x)$ is invariant under the action of the
vector field $X$ in $M$ if and only if $ \^Q (x) := Q [x,f(x)] = 0
$ (with $Q$ the characteristic of $X$).

We consider the equation $E_0 := Q = 0$ and all of its
differential consequences $E_J := D_J Q = 0$ for $|J| < k$; this
identifies the invariant manifold $\I_X \ss J^{(k)} M$. Passing to
the evolutionary representative $X_Q = Q (\pa / \pa u)$ of $X$, it
is obvious that $X_Q$ and its prolongations vanish on $\I_X$. The
$X$-invariant solutions to $\De$ will be the solutions to the
system $\De_{(X)}$ made of $\De$ and of the invariance condition:
$$ \cases{
F (x,u^{(k)}) = 0 & \cr D_J Q = 0 & ($|J|=0,...,k-1$) \ . \cr}
\eqno(15)$$
We denote the solution manifold to this system as
$\S_X \ss \I_X \ss J^{(k)} M$. The invariance of $\S_X$, as
discussed by Olver \cite{Olv1}, guarantees that the method
recalled above is justified.

Recall now that the prolongations of $X$ and $X_Q$ satisfy
$$ X^{(k)} \ = \ X_Q^{(k)} \, + \, \xi^i D_i^{(k)} \ . \eqno(16) $$

\medskip\noindent
{\bf Lemma 6.} {\it The (standard) prolongation $X^{(k)}$ of $X$
reduces to $\xi^i D_i$ on $\I_X$, and is tangent to $\S_X$.}

\medskip\noindent
{\bf Proof.} The field $X^{(k)}_Q$ vanishes on $\I_X$ because of
the equations $E_J$, and the $D_i$ are symmetries of any system,
as the differential consequences of any equation of the system are
satisfied by solutions to the system. By (16), this proves the
claim. \EOP

\subsection{Reduction of PDEs under $\mu$-symmetries}

In the case of $\mu$-symmetries of PDEs, we can proceed exactly in
the same way as for standard symmetries in order to determine
$G$-invariant solutions.

Note that the step consisting in the introduction of
symmetry-adapted coordinates is exactly the same; the difference
lies of course in the step connected to the prolongation
structure.

We describe here how the standard symmetry reduction method is
formulated to deal with $\mu$-symmetries. We suppose that $X$ is a
$\mu$-symmetry of $\De$, acting transversally for the fibration
$(M,\pi,B)$, and denote the $\mu$-prolongation of $X$ as $Y \in \X
[J^{(k)} M]$.

First of all we pass to symmetry-adapted coordinates $(y,v;\s)$ in
$M$, as in the standard case. We retain the notation introduced in subsection 1. We further proceed as there, i.e. use the chain rule to express $x$-derivatives of the $u$ as $\s$ and $y$-derivatives of the $v$. Using these, we can finally write $\De$ in terms of the
$(y,v;\s)$ coordinates and their derivatives.

Again, looking for $X$-invariant solutions means supplementing the
equation with the side condition $\pa v / \pa \s = 0$, or with
the conditions $D_J Q = 0$, see eq.(15) above, in the original
coordinates.

Now the point is that {\it if} the equation thus obtained is
independent of $\s$, we have indeed obtained a symmetry reduction
of the original equation. In this case solutions $v = \Phi (y)$ to
the reduced equation can be written in terms of the $(x,u)$
coordinates as $ \zeta (x,u) = \Phi [ \eta (x,u) ] $ and yield
implicitly $u = f (x)$, the corresponding $X$-invariant solution
to the original equation.

However, the vector field $Y$ is not the ordinary prolongation of
$X$, and thus we are not {\it apriori} guaranteed it leaves $\S_X$
or $\I_X$ invariant. Thus, in order to justify the method sketched
above -- i.e. in order to prove that the standard PDE reduction
method still applies in the case of $\mu$-symmetries -- we have to
prove the following theorem 4. Note that the only difference with
respect to the standard case will be that it is the vector field
$Y$, and not the ordinary prolongation $X^{(k)}$ of $X$, to be
tangent to the solution manifold of $\De$ in $J^{(k)} M$.

\medskip\noindent
{\bf Theorem 4.} {\it Let $\De$ be a scalar PDE of order $k$ for
$u = u(x^1,...,x^p)$. Let $X = \xi^i (\pa / \pa x^i) + \phi (\pa /
\pa u)$ be a vector field on $M$, with characteristic $Q := \phi -
u_i \xi^i$, and let $Y$ be the $\mu$-prolongation of order $k$ of
$X$. If $X$ is a $\mu$-symmetry for $\De$, then $Y : \S_X \to \T
\S_X$, where $\S_X \ss J^{(k)} M$ is the solution manifold for the
system $\De_X$ made of $\De$ and of $E_J := D_J Q = 0$ for all $J$
with $|J|=0,...,k-1$.}

\medskip\noindent
{\bf Proof.} Recall that $\S_X$ is the intersection of the
solution manifold $\S_0$ to $\De$ with the $X$-invariant set
$\I_X$ (see remark 3 above, or \cite{Win}). The former is
$Y$-invariant by assumption, as $X$ is a $\mu$-symmetry of $\De$;
the $Y$-invariance of $\I_X$ is guaranteed by lemma 5 above.
Therefore the proof for the standard case \cite{Olv1} extends to
the present setting.  \EOP

\medskip\noindent
{\bf Remark 8.} The property $Y : \I_X \to \T \I_X$ can be shown
in a alternative way without resorting to comparison with the
standard case, i.e. using the geometrical characterization of
$\mu$-prolonged vector fields, as follows.

Denote by $\I^{(m)}_X \ss J^{(k)} M$ the set of points identified
by $E_J$ for $|J| \le m$. We first show that if $\I^{(m)}_X$ is
invariant under $Y$, then $\I^{(m+1)}_X$ is also $Y$-invariant
(for $m=0,...,k-2$). Note that $Y$-invariance of $\I_X^{(m)}$
means that for all $|J| \le m$ there are functions $\b^K$ such
that $Y (D_J Q) = \sum_{|K|=0}^m \b^K D_K Q$.

We have $Y [D_i (D_J Q)] = [Y,D_i] (D_J Q) - D_i (Y (D_J Q))$;
from the corollary to lemma 4 this reads $ \la_i Y (D_J Q) + h_i^s
D_s (D_J Q) - D_i (Y (D_J Q)) + V (D_J Q) $, with $V =
\sum_{|K|=k} \ell^K (\pa / \pa u_K)$. The first term is in
$\I_X^{(m)}$ by hypothesis, while the second and third ones are by
definition in $\I_X^{(m+1)}$. The last term vanishes since $D_J Q$
does not contain $u$ derivatives of order greater than $m+1$, and
$m \le k-2$.

The proof of $Y$-invariance of $\I_X$ is hence reduced to proving
$Y$-invariance of $\I_X^{(0)}$, i.e. of the manifold identified by
$Q = 0$; as for $X$ a Lie-point vector field $Q$ depends only on
first order derivatives, it suffices to consider the first
$\mu$-prolongation of $X$, which is just $X^{(1)} + \la_i Q
\pa_{u_i}$. It is well known that $Q = 0$ is invariant under the
ordinary prolongation  $X^{(1)}$, and of course the other term
vanishes on $Q=0$.

This proves $Y$-invariance of $\I^{(0)}_X$ and hence, by the
recursive argument given above, of all the $\I^{(m)}_X$ with
$m=0,1,...,k-1$.

The recursive property considered here can be seen as a
counterpart in the PDE case to the recursive property discussed by
Pucci and Saccomandi as characterizing the $\la$-prolongations as
telescopic vector fields in the ODE case \cite{CGM}. \EOR

\section{Systems of PDEs}

In this section we extend $\mu$-prolongations to the case of $q
>1$ dependent variables.  We will assume that the dependent
variables $u$ take value in the vector space $U = \R^q$, and $B = \R^p$.

It will be natural in this context to consider differential forms
taking values in the space $\G = g\ell (q)$, the Lie algebra of
the group $G = GL (q)$. Thus we will deal with {\it matrix-valued
differential forms} (or more generally Lie-algebra valued
differential forms), see e.g. \cite{Str}. The form $\mu$ will now
be written in local coordinates as
$$ \mu \ := \ (\La_i)^a_b \, \d x^i \eqno(17) $$
where $\La_i : J^{(1)} M \to \G $ are smooth $q$-dimensional real
matrix functions. Note that remark 1 applies also to this case.

\subsection{$\mu$-prolongations in vector framework}

In this case we generalize condition (6) to the following (19); we
will then define $\mu$-prolongation in the same way as in the
scalar case.

In the vector case, we see the contact structure $\Theta$ (we use a
different symbol than in the scalar case to emphasize we deal with
vector-valued forms) as spanned by vector-valued one-forms
$\vth_\J = (\vth_\J^1 , ... , \vth_\J^q ) \in \R^q \otimes \La^1
(M)$, where
$$ \vth^a_\J \ = \ \d u^a_\J \, - \, u^a_{\J,m} \d x^m \ . \eqno(18) $$

\medskip\noindent {\bf Definition 5'.} We say that $Y$ {\bf
$\mu$-preserves the contact structure $\Theta$}, with $\mu$ given
by (17), if for any vector-valued contact forms $\vth \in \Theta$,
there is a vector-valued contact forms $\^\vth \in \Theta$ such
that
$$ \L_Y (\vth^a ) \ + \( Y \interno \[ (\La_i)^a_b \vth^b \] \) \
\d x^i \ = \ \^\vth^a \ . \eqno(19) $$ \EOD

\medskip\noindent {\bf Definition 6'.} A vector field $Y$ in $J^{(k)}
M$ which projects to $X$ in $M$ and which $\mu$-preserves the
contact structure is said to be the {\bf $\mu$-prolongation} of
order $k$, or the $k$-th $\mu$-prolongation, of $X$.  \EOD
\medskip

In order to discuss vector fields in $J^{(k)} M$ which are
$\mu$-prolongations of vector fields in $M$, it will be convenient
to agree on a general notation. That is, we write a general vector
field in $J^{(k)} M$ in the form
$$ Y \ = \ \xi^i {\pa \over \pa x^i} \ + \
\Psi^a_\J {\pa \over \pa u^a_\J} \ . \eqno(20) $$

\medskip\noindent {\bf Theorem 5.} {\it The vector field $Y$
$\mu$-preserves the standard contact structure $\Theta$ if and only if
its coefficients satisfy the {\bf vector $\mu$-prolongation formula}
$$ \Psi^a_{J,i} \ = \ \[ \de^a_b D_i \, + \, (\La_i)^a_b \] \,
\Psi^b_J \ - \ u^b_{J,k} \, \[ \de^a_b D_i \, + \, (\La_i)^a_b \]
\, \xi^k \ .  \eqno(21) $$}

\medskip\noindent
{\bf Proof.} This follows easily by a computation analogous to
that in the proof of theorem 1.  \EOP

\medskip\noindent {\bf Theorem 6.} {\it Let $X = \xi^i (\pa / \pa x^i)
+ \phi^a (\pa / \pa u^a)$ be a vector field in $M$.  Let $\mu =
(\La_i)^a_b \d x^i$ be a $\G$-valued semibasic one-form.  Then the
coefficients $\Psi^a_J$ of the $\mu$-prolongation $Y$ of $X$ are
expressed in terms of the coefficients $\wbar{\Psi}^a_J$ of the
ordinary prolongation of the same vector field $X$ as $ \Psi^a_J =
\wbar{\Psi}^a_J + F^a_J$ where the difference terms $F^a_J$
satisfy the recursion relation (with $F^a_0 = 0$) $$ F^a_{J,i} \ =
\ \[ \de^a_b D_i \, + \, (\La_i)^a_b
\] \, F^b_J \ + \ (\La_i)^a_b \, D_J Q^b \ . \eqno(22)
$$
}

\medskip\noindent
{\bf Proof.} Follow the scheme used in the proof of theorem 3. \EOP \medskip

Similarly to what happens for the $\la_i$ in the scalar case, see
remark 3, the (matrix) coefficients $\La_i$ of the form $\mu$ are
not completely arbitrary, as they must satisfy some compatibility
condition. It is convenient to define the (matrix) operators $
\nabla_i := I D_i + \La_i$.

\medskip\noindent {\bf Theorem 7.} {\it The compatibility condition
for the matrix coefficients $\La_i$ of the $\G$-valued form $\mu =
(\La_i)^a_b \d x^i$ reads
$$ D_i \, \La_j \, - \, D_j \, \La_i \ + \
\[ \, \La_i \, , \, \La_j \, \] \ = \ 0  \eqno(23) $$
for all $i,j = 1,...,p$. This is equivalent to $ \[ \, \nabla_i \,
, \, \nabla_j \, \] \ = \ 0 $.}

\medskip\noindent
{\bf Proof.} We will use the shorthand notation $(\nabla_i)^a_b =
[\de^a_b D_i + (\La_i)^a_b]$. With this, the vector
$\mu$-prolongation formula (21) reads
$$ \Psi^a_{J,k} \ = \ (\nabla_k )^a_b \Psi^b_J - u^b_{J,m} (\nabla_k)^a_b \xi^m \ ; $$
applying this twice, we get
$$ \Psi^a_{J,k,i} \ = \ \[ (\nabla_i \nabla_k )^a_b \Psi^b_J - u^b_{J,m} (\nabla_i \nabla_k)^a_b \xi^m \] \, - \, \[ u^b_{J,i,m} (\nabla_k)^a_b + u^b_{J,k,m} (\nabla_i)^a_b \] \xi^m \ ; $$
note the second square bracket is symmetric in the indices $i,k$. Thus
$$ \( \Psi^a_{J,k,i} \, - \, \Psi^a_{J,i,k} \)\ = \ [ \nabla_i ,  \nabla_k ]^a_b \Psi^b_J - u^b_{J,m} [\nabla_i , \nabla_k ]^a_b \xi^m \ . $$
As for the commutator $[\nabla_i , \nabla_k ]$, this is easily computed to be
$$ [\nabla_i , \nabla_k ] \ = \ D_i \La_k \, - \, D_k \La_i \, + \, [\La_i , \La_k ] \ , $$
i.e. the expression given in the statement, see (23). \EOP

\medskip\noindent
{\bf Remark 9.} One could consider matrices $\La_i$ belonging to a
gauged Lie algebra. By this we mean that $ \La_i  =  \la_i^k
(x,u^{(1)} ) \, L_k $, with $\la_i : J^{(1)} M \to \R$ smooth
functions and where the $L_k$ ($k=1,...,r$) are generators of a
(matrix) Lie algebra $\G$, so that $[L_i , L_j ] = c_{ij}^k L_k$.
In this case the compatibility condition reads $ \[ (D_i \la_j^k -
D_j \la_i^k ) + c_{ab}^k \la_i^a \la_j^b \] \ L_k = 0$; the term
in square brackets must vanish for each $k$. \EOR
\medskip

\subsection{$\mu$-symmetries and reduction of PDE systems}

We define $\mu$-symmetries as in the scalar case; that is, $X$ is a $\mu$-symmetry of a given PDEs system if its $\mu$-prolongation is tangent to the solution manifold of the system. For scalar equations,  $\mu$-symmetries can be used to obtain invariant solutions; the same holds for the vector case.

We will assume without further mention that $X$ satisfy the {\bf
transversality condition} in the bundle $(M,\pi,B)$ \cite{Olv1};
see remark 7.

\medskip\noindent {\bf Definition 7'.} Let $(M,\pi,B)$ be a vector
bundle over the $p$-dimensional manifold $B$, with fiber $\pi^{-1}
(x) = U = \R^q$.  Let $\De = \{ \De_1 , ... , \De_r \} $ be a
system of PDEs of order $n$ for $u^a = u^a (x)$, $a = 1,...,q$, $x
= (x^1,...,x^p) \in B$, with solution manifold $S_\De \ss J^{(n)}
M$.  Let $X$ be a vector field in $M$, and $\mu$ a $g \ell
(q)$-valued semibasic one-form on $M$ satisfying the compatibility
condition (23). Let $Y$ be the $\mu$-prolongation of order $n$ of
$X$. If $Y : S_\De \to \T S_\De$, we say that $X$ is a {\bf
$\mu$-symmetry} of $\De$.
\medskip

In the case of scalar equations, the possibility of using
$\mu$-symmetries to perform symmetry reduction relied ultimately
on two facts: (i) the space $\I_X$ of $X$-invariant functions is
$Y$-invariant for $Y$ a $\mu$-prolongations of $X$; (ii) the
standard and the $\mu$-prolongations of $X$ coincide in $\I_X$.
This entails that the results valid for reduction of an equation
$\De$ on $\I_X$ under standard symmetries extend to the case of
$\mu$-symmetries. The same holds in the case of PDEs systems.

Let us first recall that if $X = \xi^i (\pa / \pa x^i ) + \phi^a
(\pa / \pa u^a)$ is a vector field in $M$, we denote by $Q^a :=
\phi^a - u^a_i \xi^i$ its characteristic vector. Then the {\bf
$X$-invariant manifold} in $J^{(n)} M$ is the subset $\I_X \ss
J^{(n)} M$ identified by $D_J Q^a = 0$ for all $a=1,...,q$ and all
multiindices $J$ with $0 \le |J| \le n-1$.

\medskip\noindent
{\bf Theorem 8.} {\it In the hypotheses of theorem 6, let $Y$ be the
$\mu$-prolongation of the vector field $X$. Then $Y$ coincides
with the standard prolongation of the same vector field $X$ on
$\I_X$.}

\medskip\noindent {\bf Proof.} This follows from theorem 6 and the
definition of $\I_X$.  Indeed, write $\Psi^a_J$ in the form
$\Psi^a_J = \wbar{\Psi}^a_J + F^a_J$, see theorem 6, and suppose
that for $|J|=k$ the difference term $F^a_J$ is written as a
combination of the $D_J Q^b$, i.e. $F^a_J = (\Ga^J)^a_b D_J Q^b$.
Then from (22) we have
$$ F^a_{J,i} \ = \
\de^a_b [ D_i (\Ga^J)^b_c ] (D_J Q^c) + (\La_i)^a_b \[ (\Ga^J)^b_c
(D_J Q^c) + D_J Q^b \] \ ; $$ this is again a combination of terms
of the form $D_J Q^b$.  Thus if the $F^a_J$ vanish on $\I_X$ for
$|J| = k$, the $F^a_J$ with $|J| \ge k$ also vanish on $\I_X$.

Note that $F^a_i = (\La_i)^a_b Q^b$, so that the condition is
satisfied for $|J|=1$, and the proof of the theorem follows by the
recursive computation above.  \EOP \medskip

\section{Examples}

In the examples below we will consider PDEs in two independent
variables, $(x,t)$. In this case we will also write $X = \xi \pa_x
+ \tau \pa_t + \phi \pa_u$, and $\mu = \a \d x + \b \d t$.

\subsection{$\mu$-symmetries of given equations}

In order to determine $\mu$-symmetries of a given PDE $\De$ of
order $n$, one can proceed in the same way as for ordinary
symmetries. That is, consider a generic vector field $X$ acting in
$M$, and its $\mu$-prolongation $Y$ of order $n$ for a generic
$\mu = \la_i \d x^i$, acting in $J^{(n)} M$. One then applies $Y$
to $\De$, and restricts the obtained expression to the solution
manifold $S_\De \ss J^{(n)} M$. The equation $\De_*$ resulting by
requiring this is zero is the determining equation for
$\mu$-symmetries of $\De$; this is an equation for $\xi$, $\tau$,
$\phi$ and $\la_i$, and as such is nonlinear.

If we require $\la_i$ are a function on $J^{(k)} M$, all the
dependencies on $u_J$ with $|J| > k$ will be explicit, and one
obtains a system of determining equations. This system (or the
equation $\De_*$) should be complemented with the compatibility
conditions between the $\la_i$.

If we determine apriori the form $\mu$, we are left with a system
of linear equations for $\xi$, $\tau$, $\phi$; similarly, if we
fix a vector field $X$ and try to find the $\mu$ for which it is a
$\mu$-symmetry of the given equation $\De$, we have a system of
quasilinear equations for the $\la_i$.

\subsubsection*{The heat equation}

Let us first consider the heat equation
$$ u_t \ = \ u_{xx} \ ; $$
we will use the ansatz $\mu = \la_i \d x^i$ (here $x^1 = x , x^2 =
t$),
$$ \la_i \ = \ D_i \, P(x,y,u) \ ; \eqno(24) $$
this guarantees that the compatibility condition (10) is satisfied
everywhere (not just on $S_\De$).

Proceeding as mentioned above, we obtain the determining equations
for $\mu$-symmetries of the heat equation [under the ansatz (24)];
these result to be
$$ \begin{array}{l}
2 P_u \tau + 2 \tau_u = 0 \ , \\
2 P_x \tau + 2 \tau_x = 0 \ , \\
P_u^2 \xi + P_{uu} \xi + 2 P_u \xi_u + \xi_{uu} = 0 \ , \\
P_u^2 \tau + P_{uu} \tau + 2  P_u \tau_u + \tau_{uu} = 0 \ , \\
- \phi_{xx} + \phi_y - 2 \phi_x P_x - \phi P_x^2 - \phi P_{xx} + \phi  P_y = 0 \ , \\
P_x^2 \tau + P_{xx} \tau - P_y \tau + 2 P_x \tau_x + tau_{xx} - \tau_y + 2 P_x \xi + 2 \xi_x = 0 \ , \\
2 P_u P_x \tau + 2 P_{xu} \tau + 2 P_x \tau_u + 2 P_u \tau_x + 2 \tau_{xu} + 2 P_u \xi + 2 \xi_u = 0 \ , \\
-\phi_{uu} - 2 \phi_u P_u - \phi P_u^2 - \phi P_{uu} + 2 P_u P_x
\xi + 2 P_{xu} \xi + 2 P_x \xi_u + 2 P_u \xi_x + 2 \xi_{xu} = 0 \ , \\
-2 \phi_{xu} - 2 \phi_x  P_u - 2 \phi_u P_x - 2 \phi  P_u P_x -
2 \phi P_{xu} + P_x^2 \xi + P_{xx} \xi - P_y \xi + \\
\ \ \ + 2 P_x \xi_x + \xi_{xx} - \xi_y = 0 \ .\end{array} $$
After
some (lengthy but completely standard) computations, we obtain
that the more general solution to these is given by
$$ \begin{array}{rl}
\xi  (x,t,u) \ =& \ e^{- P} \ \[ c_1 + c_2 t + (c_3/2) x + (c_4/2) x t  \] \ , \\
\tau (x,t,u) \ =& \ e^{- P} \ \[ c_5 + c_3 t + (c_4/2) t^2 \] \ , \\
\phi (x,t,u) \ =& \ e^{- P} \ \[ \z (x,t) \, + \, \( - (c_2/2) x -
(c_4 / 8) (x^2 - 2 x t ) + c_6 \) u \] \ , \end{array} $$ where
$c_i$ are arbitrary constants, and $\z (x,t)$ is an arbitrary
function satisfying $\z_t = \z_{xx}$. Thus, we just obtain the
standard symmetries of the heat equation \cite{Gae,Olv1,Ste}, with
the factor $\exp [- P (x,y,u)]$; this is no accident, but follows
from the ansatz (24), see remark 5 (see also \cite{CGM}). The
characteristic $Q := \phi - \xi u_x - \tau u_t$ will be the same
as for standard symmetries (with a factor $e^{-P}$), and the
symmetry reduced equations will give nothing new.

\subsubsection*{The Euler equation}

Let us consider the Euler equation
$$ u_t \ + \ u \, u_x \ = \ 0 \ ; $$
we will write as usual $X = \xi \pa_x + \tau \pa_t + \phi \pa_u$,
and $\mu = \a \d x + \b \d y$.

The condition for $X$ to be a $\mu$-symmetry for the Euler
equation is that
$$ \begin{array}{l}
\phi u_x + \a u \phi + \b \phi + u^2 u_x \a \tau + u u_x \b \tau - u u_x \a \xi -
u_x \b \xi + \\
\ \ \ \ + \phi_t + u u_x \tau_t - u_x \xi_t + u \phi_x + u^2 u_x
\tau_x - u u_x \xi_x \ = \ 0 \ .
\end{array} \eqno(25)
$$
This should be complemented with the requirement that $D_x \b =
D_t \a$ when $u_t + u u_x = 0$. With the ansatz
$$ \a = \a (x,t,u) \ , \ \b = \b (x,t,u) \ , \eqno(26) $$
(note (24) cannot be verified in this case if $\a$ and $\b$ do
actually depend on $u$) the dependence of the equations above in
$u_x$ is explicit, and (25) splits into two equations:
$$  \begin{array}{l}
(\a u + \b ) \phi + \phi_t + u \phi_x \ = \ 0 \ ; \\
\phi + (\a u^2 + \b u ) \tau - (\a u + \b ) \xi + u \tau_t - \xi_t
+ u^2 \tau_x - u \xi_x \ = \ 0 \ . \end{array} $$ These are again
nonlinear equations for the functions $(\a,\b,\xi,\tau,\phi)$. A
special solution is provided e.g. by
$$ \begin{array}{l}
\a = u \ , \ \b = - u^2 / 2 \ ; \\
\xi = 0 \ , \ \tau =  [B(u) - A(u) t / u] \ \exp[- (u^2/2) t ] \ ,
\ \phi = A(u) \ \exp[- (u^2/2) t ] \ . \end{array} $$ Note that
for this $\mu$, the compatibility conditions $D_t \a = D_x \b$ is
satisfied only on the solution manifold $S_\De$, see remark 4.

This $\mu$-symmetry corresponds to a nonlocal ordinary symmetry
$Z$ of exponential type, see remark 5. We have in facts
$$ Z \ = \  e^{\int (u \d x - (u^2 /2 ) \d t ) } \ X \ .
$$

\subsection{Equations with given $\mu$-symmetries}

We can also consider the opposite question, i.e. given a vector
field $X$ and a form $\mu = \la_i \d x^i$ satisfying (10),
determine the equations of a given order $n$ which admit $X$ as a
$\mu$-symmetry with the given $\mu$.

To solve this problem, we have to consider the $\mu$-prolongation
$Y$ of $X$ to $J^{(n)} M$, and solve the characteristic equation
for it. In this way we obtain the differential invariants for $Y$,
and any equations which is written in terms of these will admit
$X$ as a (strong) $\mu$-symmetry.

\subsubsection*{Example 1}

As a first example, to be dealt with in detail, we will consider
$\mu$-prolongations of the scaling vector field
$$ X = x \pa_x + 2 t \pa_t + u \pa_u \ . $$
The invariant coordinates $(y,v)$ and the parametric coordinate
$\s$ in $M = \{(x,t,u)\}$ can be chosen as $ \s = x \ , \ y =
x^2/t \ , \ v = u/x $; the corresponding inverse change of
variables is $ x = \s \ , \ t = \s^2 / y \ , \ u = \s v $.

It follows easily that in the symmetry-adapted coordinates, $ X =
\s \pa_\s$; hence the function $v = v(\s,y)$ is $X$-invariant if
and only if $v_\s = 0$, as required by the general method (indeed,
by the very definition of symmetry-adapted coordinates).

Applying the procedure described in sect.4, we have
$$ u_x  =  v + 2 y v_y + \s v_\s \ ; \
u_t = - (y^2 /\s) v_y \ . $$ The above can be inverted to give
$v_\s = (1/x) [ u_x + 2(t/x) u_t - u/x ] = - Q/x^2$, $v_y = - [t^2
/ x^3] u_t$. Similarly, at second order we get
$$ \begin{array}{l}
u_{xx} = 2 v_\s + 6 (y/\s) v_y + \s v_{\s \s} + 4 y v_{\s y} + 4
(y^2 / \s ) v_{yy} \ , \\
u_{xt} = - 3 (y^2 / \s^2 ) v_y - (y^2 / \s ) v_{\s y} - 2 (y^3 /
\s^2 ) v_{yy} \ , \\
u_{tt} = 2 (y^3 / \s^3) v_y + (y^4 / \s^3 ) v_{yy} \ .
\end{array} $$

We will now consider the simplest nontrivial choice for $\mu$,
i.e. $\mu = \la \d x$, with $\la$ a real constant.

The second $\mu$-prolongations can be written in the form (20)
(with $q=1$), and general explicit expressions for the
coefficients are obtained using either (12) or (14). With our
choice for $\mu$, one gets
$$ \begin{array}{l}
\Psi^x = \la \, (u - x u_x - 2 t u_t) \ , \ \Psi^t = - u_t \ , \\
\Psi^{xx} = - u_{xx} - 2 \la (x u_{xx} + 2 t u_{xt}) + \la^2 (u - x u_x - 2 t u_t) \ , \\
\Psi^{xt} = - 2 u_{xt} - \la (x u_{xt} + 2 t u_{tt} + u_t ) \ , \
\Psi^{tt} = - 3 u_{tt} \ . \end{array} $$

By the method of characteristics, we obtain with standard
computations that the invariants of $Y$ in $J^{(2)} M$ are given
by:
$$ \begin{array}{l}
y := (x^2 / t ) \ , \ \ \ v := (u / x) \ ; \\
\z_1 := x u_t \ , \ \ \ \z_2 := (u/x - 2 t u_t/x - u_x ) e^{\la x} \\
\eta_1 := x t u_{tt} \ , \ \ \ \eta_2 := \(x u_t + 2 x t u_{tt} + x^2 u_{xt} \) e^{\la x} \ , \\
\eta_3 := ( 1/x) \[ (1 - \la x) (u - x u_x) + 2 \la x t u_t + x^2
u_{xx} + 4 x t u_{xt} + 4 t^2 u_{tt} \] e^{2 \la x} \ .
\end{array}
$$

Thus, any equation $ \De := \ F \[ y,v,\z_1 , \z_2 , \eta_1 ,
\eta_2 , \eta_3 \] = 0$, with $F$ an arbitrary smooth function of
its arguments, admits the vector field $X$ given above as a
$\mu$-symmetry, with $\mu = \la \d x$. Moreover, for $(\pa F / \pa
\z_2)^2 + (\pa F / \pa \eta_2)^2 + (\pa F / \pa \eta_3)^2 \not=
0$, $X$ is not an ordinary symmetry of $\De$.

Finally, it is easily seen by restricting the functions given
above to $\I_X$ that the $\mu$-symmetry reduced equation,
providing $X$-invariant solutions to $\De$, is given by
$$ H [ y,v,\z_1,\eta_1] \ := \ F \[ y,v,\z_1 , 0 , \eta_1 , 0 , 0 \] \ = \ 0 \ . $$

Let us discuss a completely concrete example. Consider the
equation $\De := \eta_1 - \z_2 = 0$. This is written as
$$ x t \, u_{tt} \ + \ \( u_x + (2 t/x) u_t - u/x \) \, e^{\la x} \ = \ 0 $$
in the original coordinates; in the adapted ones it reads
$$ y^3 v_{yy} + 2 y^2 v_y + \s e^{\la \s} v_\s \ = \ 0 \ . $$
The corresponding reduced equation is $ y^2  [ y v_{yy} + 2 v_y ]
= 0$; the general solution to this is $ v (y) = c_1 + c_2 / y  $,
where $c_i$ are real constants. Going back to the original
coordinates, the corresponding solutions are $ u (x,t) =  (c_1
x^2 +  c_2  t)/x$.

\subsubsection*{Example 2.}

Consider the same $X$ as above, with $\mu = - (1/t) \d t$. In this
case the $Y$-invariant functions are spanned by
$$ \begin{array}{l}
\z_1 = u_x \ ; \ \ \ \z_2 = u_t / x - (u/x - u_x)/ (2 t ) \ ; \\
\eta_1 = x u_{xx} \ , \ \ \ \eta_2 = u_{xt} + x u_{xx}/ (2 t) \ , \\
\eta_3 = (x u_{xx} -3 u_x)/(4 t^2) + u_{xt}/t  + u_{tt}/x  -
u_t/(x t) + 3 u/(4 x t^2) \ . \end{array} $$ Any equation given by
$F[y,v,\z_1 , \z_2, \eta_1 , \eta_2, \eta_3] = 0$ will admit $X$
as a (strong) $\mu$-symmetry, and if $(\pa F / \pa \z_2)^2 + (\pa
F / \pa \eta_2)^2 + (\pa F / \pa \eta_3)^2  \not=0$, $X$ is not an
ordinary symmetry.

The restriction of the invariant functions to $\I_X$ yields $ \z_1
= u_x$, $\z_2 = 0$, and $\eta_1 = x u_{xx}$, $\eta_2 = \eta_3 =
0$;  hence the reduced equation will be simply
$$ H [y,v,\z_1,\eta_1] \ := \ F[y,v;\z_1 , 0 ; \eta_1 ,0,0] \ = \
0 \ . $$

\subsubsection*{Example 3.}

Any equation of the form $ F[ y , v , \z_1,\z_2 , \eta_1 , \eta_2
, \eta_3 ] =  0 $, where $F$ is a smooth function of its arguments
and we have defined
$$ \begin{array}{l}
y = \sqrt{x^2 + t^2} \ , \ \ \ v = u \ , \\
\z_1 = u_x / x \ , \ \ \ \z_2 = u_t - (t/x) u_x \ , \\
\eta_1 = (y^2 / x^3 ) (x u_{xx} - u_x ) \ , \ \ \
\eta_2 = (y / x^3) (x t u_{xx} - x^2 u_{xt} - t u_x ) \ , \\
\eta_3 = u_{tt} + (t/x^3) (x t u_{xx} - 2 x^2 u_{xt} - t u_x ) \ ,
\end{array} $$ admits the rotation vector field
$$ X \ = \ x \pa_t - t \pa_x $$
as a $\mu$-symmetry, with $\mu = - (1/x) \d x$.

The functions $ y, v$ provide invariants in $M$, and we can select
$\s = {\rm arctg} (t/x)$; the inverse change of coordinates is
given by $ x = y \cos (\s)$, $t = y \sin (\s)$, $u = v$. The
vector field $X$ is then expressed as $X = \pa_\s$

The invariant subset $\I_X$ is in this case identified by $ u_t =
(t/x) u_x$, $u_{xt} = (x u_{xx} - u_x ) (t/x^2)$, $u_{tt} = [x t^2
u_{xx} + (x^2 - t^2 ) u_x ] / x^3$. The restriction of the
invariant functions to $\I_X$ yields $ \z_1 = u_x / x$, $\z_2 =
0$, $\eta_1 = (r^2/x^3) (x u_{xx} - u_x)$, $\eta_2 = 0$, $\eta_3 =
u_x/x$ (note $\eta_3 = \z_1$); hence the reduced equation will be
simply
$$ H [y,v,\z_1,\eta_1] \ := \ F[y,v;\z_1 , 0 ; \eta_1 ,0,\z_1 ] \ = \
0 \ . $$

\subsubsection*{Example 4.}

In previous examples the functions $\la$ and $\nu$ in $\mu = \la
\d x + \nu \d t$ were always depending only on $x$ and $t$; in
this last example they will depend on first order derivatives of
the $u$.

Any equation of the form $F ( y,v; \zeta_1,\zeta_2;
\eta_1,\eta_2,\eta_3 ) =  0 $ with $F$ a smooth function of its
arguments, which are
$$ \begin{array}{l}
y \ = \ t \ , \ \ \ v \ = \ u/x \ ; \\
\z_1 = u + \log[1 - u/(x u_x)] \ , \ \ \
\z_2 = -(u u_t)/(x^2 u_x) \ ; \\
\eta_1 \ = \ - e^{2 u} \, \[ u  / (x^2 u_x^3) \] \,
\[ u^2 (u_x^2 - u_{xx} ) + x u_x^3 - (1 + x u_x) u u_x^2 \] \ , \\
\eta_2 \ = \ e^u \, \[ u^2  / (x^3 u_x^3) \] \, \[ x u_x u_{xt} - (u_x + x u_{xx} ) u_t \] \ , \\
\eta_3 \ = \ \[ u / (x^3 u_x^3 ) \] \, \[ x u_x^2 u_{tt} - 2 x u_x
u_t u_{xt} + (2 u_x + x u_{xx} ) u_t^2 \] \ ,  \end{array} $$
admits the scaling vector field
$$ X \ = x \,  \pa_x \ + \ u \, \pa_u $$
as a (strong) $\mu$-symmetry, with $\mu = u_x \d x + u_t \d t $.

The invariant set $\I_X$ is identified by $ u_x = u/x$, $u_{xt} =
u_t / x$, $u_{xx} = 0$. Restriction of the invariant functions to
$\I_X$ yields $ \z_1 = \z_1^0 := u/x = v$\footnote{We stress that
this expression for $\z_1$ is not obtained by a direct
substitution: indeed now $Q=0$ means $u_x = u/x$; the general
expression for $\z_1$ given above becomes singular, but the
expression for $\Psi^x$ guarantees that $u_x$ is constant, and
actually equal to $u/x = v$ on $Q=0$.}, $\z_2 = \z_2^0 := - u_t /
x$, $\eta_1 = \eta_2 = 0$, $\eta_3 = \eta_3^0 := u_{tt} / x$.
Hence the reduced equation will be simply
$$ H [y,v,\z_2^0,\eta_3^0] \ := \ F[y,v;v,\z_2^0;0,0,\eta_3^0] \ = \ 0 \ . $$

\subsection{Systems of PDEs}

Finally, we consider systems of PDEs with a given $\mu$-symmetry,
and the corresponding reduction. We will denote independent and
dependent variables as $(x,y)$ and $(u,v)$ respectively.

\subsubsection*{Example 1}

Let us consider the scaling vector field
$$ X \ = \ x {\pa \over \pa x} + 2 y {\pa \over \pa y} + u {\pa \over \pa u} + 2 v {\pa \over \pa v} $$
and the form $ \mu \ = \ \la I \d x $ with $\la$ a real constant;
this corresponds to matrices $\La_i$ given by $ \La_{(x)} = \la I$
and $\La_{(y)} = 0$.

By applying the vector $\mu$-prolongation formula (21), or using
theorem 6 and (22), we determine the second $\mu$-prolongation $Y$
of $X$. We can then solve the characteristic equation for the flow
of $Y$ in $J^{(2)} M$, and determine a basis of $Y$-invariant
functions. Such a basis is provided by the following set of
functions:
$$ \begin{array}{l}
\rho = x^2 / y \ , \ w_1 := u/x \ , \
w_2 := v / x^2 \ ; \\
\z_1 := y u_y / x \ , \
\z_2 := (u_x + 2 y u_y / x - u/x ) e^{\la x} \ , \\
\z_3 := v_y \ , \
\z_4 := (v_x/x - 2 v/x^2 + 2 y v_y / x^2) e^{\la x} \ ; \\
\eta_1 := x^2 v_{yy} \ , \
\eta_2 := x^3 u_{yy} \ , \\
\eta_3 := (x v_{xy} + 2 y v_{yy} ) e^{\la x} \ , \\
\eta_4 := (x^2 u_{xy} + x u_y + 2 x y u_{yy} ) e^{\la x} \ , \\
\eta_5 := [ v_{xx} + 4 v / x^2 - 3 v_x / y + 4 y u_{xy} - 4 y v_y
/
x^2 + 4 y u_y / x + \\
\ \ \ \ - 4 y^2 v_{yy} / x^2 + 8 y^2 u_{yy}/x + \la ( v_x - 2 v/x
+ 2
y v_y / x ) ] e^{2 \la x} \ , \\
\eta_6 := [ -u_x + u/x + x u_{xx} + 4 y u_{xy} + 4 y^2 u_{yy}/x + \\
 \ \ \ \ + \la ( -u + x u_x + 2 y u_y) ] e^{2 \la x} \ . \end{array} \eqno(27) $$
Any (system of) second order equation of the form
$$ F^i \[ y , w_1, w_2;
\z_1,...,\z_4 ; \eta_1 , ...  , \eta_6 \] = 0 \eqno(28) $$ with
$F^i$ ($i=1,...,n$) a smooth function of its arguments, admits $X$
as a (strong) $\mu$-symmetry.

In order to consider the $\mu$-symmetry reduced equation, it
suffices to consider the restriction of the functions $\z_i ,
\eta_j$ on $\I_X$ (note that the $\mu$-prolongation and the
ordinary one coincide on $\I_X$, see lemma 5.) The manifold $\I_X$
is identified by $Q= D_x Q = D_y Q = 0$; in the present case these
mean
$$ \begin{array}{lll}
u_y = (u - x u_x) / (2 y) \ , & u_{xy} = - (x u_xx ) / (2 y) \ , &
u_{yy} = - (u_y + x u_{xy} )/(2y) \ ; \\
v_y = (2 v - x v_x) / (2y) \ , & v_{xy} = (v_x - x v_{xx} ) /(2y)
\ , & v_{yy} = - (x v_{xy} ) / (2 y) \ .  \end{array} $$

Substituting these into (27) above, we obtain the expressions for
the reduction of first and second order $Y$-invariants restricted
to $\I_X$, which are
$$ \begin{array}{l} \z_1^0 = ({u - u_x \,x})/({2\,x})
\ , \  \z_2^0 = 0 \ , \\
\z_3^0 = ({2 v - x v_x})/({2y}) \ , \ \z_4^0 = 0 \ ; \\
\eta_1^0 = ({{x^3}\,( - v_x + x v_{xx} ) })/({4 y^2}) \ , \\
\eta_2^0 = ({{x^3}\,( - u + x \,( u_x + x u_{xx} ) ) })/({4
{y^2}}) \
, \\
\eta_3^0 = 0  \ , \ \eta_4^0 = 0 \ , \ \eta_5^0 = 0 \ , \ \eta_6^0
= 0 \ . \end{array}  $$ Thus, the $X$-invariant solutions of (28)
are obtained as solution of the reduced system of equations
$$ H^i [y,w_1,w_2;\z_{1}^0 , \z_3^0, \eta_1^0 , \eta_2^0 ] \ := \ F^i \[ y , w_1, w_2; \z_1^0,0,\z_3^0,0 ; \eta_1^0 , \eta_2^0 , 0,0,0,0 \]
= 0\ . $$

\subsubsection*{Example 2}

Consider next the elementary vector field $ X = x \pa_x $, and the
form $ \mu = \La_{(x)} \d x + \La_{(y)} \d y$ corresponding to
matrices $$ \La_x \ = \ \pmatrix{0&-y\cr0&0} \ , \ \La_y = {1
\over y^2} \pmatrix{- x y & (x^2 - x) y^2 \cr 1 & x y \cr } \ .
$$

In this case, a basis for $Y$-invariant functions on $J^{(2)} M$
is provided by
$$ \begin{array}{l}
\rho = y \ , \ \ w_1 = u \ , \ \
w_2 = v \ ; \\
\z_1 = x u_x - x^2 y v_x \ , \ \
\z_2 = x v_x \ , \\
\z_3 = (1/y) [y u_y - x^2 y v_x - x ( x u_x - x^2 y v_x)] \ , \\
\z_4 = v_y - (1/y^2) \, (x u_x - x^2 y v_x) \, \log (x) \ ; \\
\eta_1 = x^2 u_{xx} - 2 x^2 y v_x - 2 x^3 y v_{xx} \ , \ \
\eta_2 = x^2 v_{xx} \ , \\
\eta_3 = x [u_{xy} - x (2 v_x + x v_{xx} + y v_{xy})] \ , \\
\eta_4 = - (1/y^2) \{ x y (x v_x + x^2 v_{xx} - y v_{xy} ) + x
[u_x +
x (u_{xx} - 3 y v_x - 2 x y v_{xx})] \, \log (x) \} \ , \\
\eta_5 = (1/y^2) [ 4 x^2 u_x + 2 x^3 u_{xx} + y (-2 x^2 u_{xy} - 4
x^3
v_x - 2 x^4 v_{xx} + \\
\ \ \ \ + y u_{yy} - 2 x^2 y v_{xy} + 2 x^3 y v_{xy})] \ ,
\\
\eta_6 = (1/y^3) \{ -2 x^2 y v_x - 2 x^3 y v_{xx} + y^3 v_{yy} + \\
\ \ \ \ + 2 x [u_x + y (x v_x + x^2 v_{xx} + x y v_{xy} - u_{xy})]
\log (x) \} \ .
\end{array} $$

Any system of second order equation of the form $$ F^i
\[ y, w_1, w_2; \z_1,...,\z_4 ; \eta_1 , ...  , \eta_6 \] = 0 \ , $$
$F^i$ a smooth function of its arguments, admits $X$ as a (strong)
$\mu$-symmetry.

The system identifying $\I_X$ is now given by $$
\begin{array}{lll} x
u_x = 0 , & u_x + x u_{xx} = 0 , & x u_{xy} = 0 , \\
x v_x = 0 , & v_x + x v_{xx} = 0 , & x v_{xy} = 0 \ ; \end{array}
$$ $\I_X$ is the linear space on which $ u_x = u_{xx} = u_{xy} =
v_x = v_{xx} = v_{xy} = 0$. Restriction of the invariant functions
given above to this space yields
$$ \begin{array}{l} \z_1^0 = \z_2^0 = 0 \ , \ \z_3^0 = u_y \ , \
\z_4^0 = v_y \ ; \\
\eta_1^0 = \eta_2^0 = \eta_3^0 = \eta_4^0 = 0 \ , \ \eta_5^0 =
u_{yy} \ , \ \eta_6^0 = v_{yy} \ .  \end{array} $$ The reduced
system is therefore
$$ H^i [y,u,v;u_y,v_y,u_{yy},v_{yy} ] \ = \ F^i
[y,u,v;0,0,u_y,v_y;0,0,0,0,u_{yy},v_{yy} ] \ = \ 0 \ . $$

\vfill\eject

\end{document}